\documentclass[%
 aip,
 amsmath,amssymb,
 reprint,%
]{revtex4-1}

\usepackage{graphicx}% Include figure files
\usepackage{dcolumn}% Align table columns on decimal point
\usepackage{bm}% bold mathhttps://www.overleaf.com/project/635a553088d537a945b934b5
\usepackage{xcolor}
\usepackage[utf8]{inputenc}
\usepackage[T1]{fontenc}
\usepackage{mathptmx}
\usepackage{float}
\usepackage{etoolbox}

\makeatletter
\def\@email#1#2{%
 \endgroup
 \patchcmd{\titleblock@produce}
  {\frontmatter@RRAPformat}
  {\frontmatter@RRAPformat{\produce@RRAP{*#1\href{mailto:#2}{#2}}}\frontmatter@RRAPformat}
  {}{}
}%

\usepackage{soul}
\definecolor{grey}{RGB}{175,175,175}
\definecolor{red}{RGB}{200,0,0}
\definecolor{green}{RGB}{150,175,150}
\definecolor{blue}{RGB}{0,0,200}
\definecolor{orange}{RGB}{255,88,0}
\definecolor{yellow}{RGB}{255,252,201}
\makeatother
\begin{document}

\preprint{AIP/123-QED}

%\title[Parameter and coupling estimation in Izhikevich neurons]{Parameter and coupling estimation in Izhikevich neurons}
\title[{Parameter and coupling estimation in small groups of Izhikevich's neurons}]{{Parameter and coupling estimation in small  {networks} of Izhikevich's neurons}}

% Force line breaks with \\
\author{R. P. Aristides}
\affiliation{Instituto de F\'{i}sica Te\'{o}rica, Universidade Estadual Paulista,  Rua Dr. 
			Bento Teobaldo Ferraz 271, Bloco II, Barra Funda, 01140-070 S\~ao Paulo, Brazil.%\\This line break forced with \textbackslash\textbackslash
}%
\affiliation{%
Departament de Fisica, Universitat Politecnica de Catalunya, St. Nebridi 22, 08222 Terrassa, Spain%\\This line break forced% with \\
}

\author{A. J. Pons}
 \email{a.pons@upc.edu}
\affiliation{%
Departament de Fisica, Universitat Politecnica de Catalunya, St. Nebridi 22, 08222 Terrassa, Spain%\\This line break forced% with \\
}%

\author{H. A. Cerdeira}%
\affiliation{Instituto de F\'{i}sica Te\'{o}rica, Universidade Estadual Paulista,  Rua Dr. 
			Bento Teobaldo Ferraz 271, Bloco II, Barra Funda, 01140-070 S\~ao Paulo, Brazil.%\\This line break forced with \textbackslash\textbackslash
}%
\author{C. Masoller}
 \email{cristina.masoller@upc.edu}
 \homepage{http://www.fisica.edu.uy/~cris/}
\affiliation{%
Departament de Fisica, Universitat Politecnica de Catalunya, St. Nebridi 22, 08222 Terrassa, Spain%\\This line break forced% with \\
}%

\author{G. Tirabassi}%
\affiliation{%
Departament de Fisica, Universitat Politecnica de Catalunya, St. Nebridi 22, 08222 Terrassa, Spain%\\This line break forced% with \\
}%

\date{\today}% It is always \today, today,

\begin{abstract}
{{
{Nowadays, experimental techniques allow scientists to have access to large amounts of data. In order to obtain reliable information from the complex systems which produce these data, appropriate analysis tools are needed}. 
The Kalman filter is a  {frequently used} technique to infer, assuming a model of the system, the parameters of the model from uncertain observations. A well-known implementation of the Kalman filter, the Unscented Kalman filter (UKF), was recently shown to be able to infer the connectivity of a set of coupled chaotic oscillators. {I}n this work, we test whether the UKF can also reconstruct the connectivity of {small groups of} coupled neurons when their links are either electrical or chemical {synapses}. {In particular, w}e consider Izhikevich neurons, and aim to infer which neurons influence each other, considering {simulated spike trains
as the experimental observations used by the UKF}. First, we  {verify} that the UKF can recover the parameters of a single neuron, even when the parameters vary in time. Second, we analyze small neural ensembles and}} demonstrate that the UKF allows inferring the connectivity between the neurons, even for heterogeneous, directed, and {temporally evolving} networks. {Our results show that time-dependent parameter and coupling estimation is possible in this nonlinearly coupled system}.
\end{abstract}

\maketitle
\begin{quotation}
The Kalman filter is a popular technique that can be employed to infer the parameters of a model given uncertain observations, and it has found applications in diverse fields. In the field of neuroscience, it has been used, for example, to estimate the parameters of neural models, and for real-time decoding of brain signals for brain-machine interfaces. However, the neural models that have been considered contain a large number of parameters, which makes a systematic exploration of the parameter space unfeasible. Here we study a neural model, the Izhikevich model, which realistically reproduces many neural states, even though it is computationally low-cost. Having a small number of parameters {and, at the same time, showing very rich dynamical regimes}, the Izhikevich model {is} an ideal candidate for a systematic exploration of the parameter space {and the} study {of} neurons coupled with different topologies. We analyze the suitability of the Kalman filter to estimate the model's parameters and we discuss its main limitations.
\end{quotation}

\section{Introduction}

One of the main challenges that neuroscience has faced for a long time is the determination of brain topology, which is morphologically diverse and complex. %The topology of the brain network is morphologically diverse~\cite{Schuz2002} and very complex~\cite{Mountcastle1997}. 
Besides, the elements that form the brain network, the neurons, are %the dynamics of the main elements which form the nodes of the brain network, the neurons, is 
also diverse and complex. 
 {Neurons show reproducible nonlinear responses to stochastic stimuli~\cite{Bryant1976}. Hence, they can be modeled as stochastic nonlinear dynamical systems.~\cite{Sterratt2011}}

%Neurons are stochastic; however, they show %~\cite{Faisal2008,Tuckwell88,Laing,Tuckwell89}. However, they
%reproducible responses to stochastic stimulus~\cite{Bryant1976}, meaning that neurons can be modeled as stochastic nonlinear dynamical systems~\cite{Sterratt2011}. %~\cite{Janson2012,Sterratt2011}.
%In fact, noise has been shown to play constructive roles in neural dynamics~\cite{Ward2009}.

Although much progress has been made on the relationship between topology and dynamics in the brain, scientists are still far from having a good understanding  {\cite{bassett2017network,battiston2020networks}}. 
Mathematical models of the whole brain or just of a tiny fraction of billions of neurons, as well as information-based data analysis techniques are powerful tools for shedding light on the above relationship \cite{Goodfellow2022}.

%Some progress has been made, also, in the understanding of functional brain networks~\cite{Bassett2017,Sporns2011} and
%efforts trying to explain emergent processes observed in the brain are appearing more often nowadays~\cite{Izhikevich2008}. So, due to the complex diversity of structures and dynamics of the brain, the understanding of its ability to encode~\cite{Averbeck2006,DayanAbbott} and process information, at different spatio-temporal scales~\cite{Steriade2001,Buzsaki2004,Buzsaki2009}, is still out of reach for the scientific community. 
%Nevertheless, several information-based techniques are becoming available to forecast brain's behavior~\cite{Ponce-Flores2020}.
 {However, a realistic estimation of the models' states and parameters is a very difficult challenge, and different approaches based on control theory have been developed~\cite{Bechhoefer2021}. A well-known method is the Kalman filter.~\cite{Kalman1,Haykin2009,Bishop2016}} 
%In many cases, a mixture of rough estimates, experimental measurements or, even, inter-species parameters values are used to feed theoretical models of neurons and networks. CITATION NEEDED
%Nevertheless, the realistic estimation of states and parameters 
%different approaches based on control theory have been developed~\cite{Bechhoefer2021}.
%~\cite{Tewari2002,Bechhoefer2021} for a long time already. 
%Statistical inference is one of them that allows to establish parameters in model descriptions which are compatible with experimental data. A frequently applied method is the Kalman filter~\cite{Kalman1,Haykin2009,Bishop2016}. 

The Kalman filter allows inferring optimal parameters of a model given uncertain observations, balancing the effects of measurement noise, disturbances, and model uncertainties, and has found applications in many fields of science and technology~\cite{Kutz2019}. In neuroscience, the Kalman filter has been used, for example, for decoding brain signals for brain-machine interfaces~\cite{bmi2009,bmi2018,bmi2021}. It has also been used to estimate the parameters of neural models~\cite{moris,hh1,hh2,hh3,hh4,Schiff2012}. However, the models that have been considered, such as the Morris–Lecar or the Hodgkin-Huxley, contain a large number of parameters that  {make a systematic exploration of the parameter space unfeasible}. Here we study the Izhikevich model~\cite{izke1}(IM) because it reproduces many important properties of biological neurons and, at the same time, has a small number of parameters and is computationally low-cost\cite{izke2}. Therefore, the Izhikevich model is an ideal candidate for a systematic exploration of the parameter space allowing a study of small ensembles of coupled neurons. 

We analyze under which conditions a  nonlinear version of the Kalman filter, the Unscented Kalman Filter (UKF)~\cite{kf2,kf4}, provides a good estimation of the IM parameters and we discuss its main limitations.
We show that the UKF is able to recover the parameters of an isolated neuron and the external current that is exciting its activity. We also show that the UKF is able to do so even in the case of time-dependent input currents. Then, we study small networks with different topologies, with both electrical and chemical couplings,  {and show that UKF is able to recover the topology of the network using observations of the dynamic variables, assuming the coupling strength, electrical or chemical, and all the internal parameters are known.}

\section{Methods}
\subsection{Model}
%\subsection{The Izhikevich Model}

The Izhikevich model (IM) was introduced by Eugene M. Izhikevich~\cite{izke1} as an alternative to more realistic but computationally expensive neuron models\cite{Izhikevich2004}. Despite its simplicity, it can be used to model a broad variety of neuron types \cite{izke2} {and dynamical regimes}. Here will focus on single Izhikevich neurons in the chaotic regime  { --- that is neurons for which the spiking dynamics is irregular, aperiodic ---} and small networks of chaotic neurons linked by electrical or chemical couplings. 

The state of an Izhikevich neuron $i$ is fully specified by two state variables. $x_i$ represents the neuron membrane potential and $y_i$ represents the membrane recovery variable accounting for the activation of the ionic currents.

Let $[x_1, y_1, \dots, x_i, y_i, \dots]^T$ be the state vector of the neurons %\comment{This array will be larger when considering small networks, maybe I would write the general form of the state vector considering the small motif here}
,  {the equations governing the system} are given by
\begin{equation} \label{izk}
\begin{split}
\dot x_i &= 0.04\,x_i^2 + 5x_i + 140 - y_i +I + E_i + C_i + \sigma_Z\xi^x_i \\
 \dot y_i &= a\,(b\,x_i - y_i) + \sigma_Z\xi^y_i
\end{split}
\end{equation}

%\comment{We are mixing notation: Why not $\xi^x_i$ and $\xi^y_i$ instead of $\xi^v_i$ and $\xi^u_i$}
with the after-spike reset condition: 

\begin{equation}
    \text{if}\quad x_i > 30, \quad\text{then}
    \begin{cases}
      x_i \rightarrow c, \\
      y_i \rightarrow y_i + d. 
    \end{cases} \ 
    \label{reset} 
\end{equation} 

$a$ is a small parameter representing the slow time-scale of $y_i$, $b$ is the coupling strength between the state variables, and the external currents are modeled by $I$.  {All parameters here, included $x$, $y$ and time are dimensionless}. The parameters $a$, $b$, $c$ and $d$ can be fitted to obtain a specific firing pattern of the neuron. The last term in Eq.\eqref{izk} represents random fluctuations and we refer to {it} as dynamic or process noise. $\xi^x_i$ and $\xi^y_i$ represent Gaussian white noises with zero mean and unity variance. $\sigma_Z$ is the noise strength and for simplicity, it is the same for $x$ and $y$. For a system of $N$ neurons, the dynamical noise can be thought as a random 2$N$-dimensional vector with zero mean and covariance matrix $\bar Q_Z=\sigma_Z^2\mathbb{I}$, where $\mathbb{I}$ is the identity matrix.

The electrical coupling between neurons is described by a system of ordinary first-order differential equations with different levels of detail that represent various degrees of physiological descriptions \cite{perkel}.
Here we consider the simplest coupling,  {namely linear diffusive coupling, $E_i$ is given by}
\begin{equation}
    E_i =  g_e \sum_{j=1}^N A^e_{ij} (x_j - x_i) \ .
    \label{eq:elec}
\end{equation}
where $g_e$ is the coupling conductance and $A^e_{ij}$ are the coefficients of the adjacency matrix: $A^e_{ij}=1$ whenever neuron $i$ is connected to neuron $j$, otherwise $A^e_{ij}=0$.

 {The coupling $C_i$ comprises inputs delivered through chemical synapses to neuron $i$ from all other neurons in the network. It is given by\cite{somers}}: 
\begin{equation}
    C_i = g_c (x_i - \mu_s) \sum_{j = 1}^N A^c_{ij} \zeta(x_j).
    \label{eq:chem}
\end{equation}
$g_c$ is the synaptic coupling strength, $\mu_s$ is the reversal potential, and the sigmoid function $\zeta(x)$ is defined as
\begin{equation} 
    \zeta(x_j) = [1 + \exp(-\epsilon(x_j-\theta))]^{-1}, \ \label{sigmoid}
\end{equation}
where $\epsilon$  {controls the} slope of the  {sigmoidal} function and $\theta$ is the synaptic firing threshold.  {This function represents the activation of the postsynaptic current when a presynaptic neuron sends an action potential, that is when $x$ becomes larger than $\theta$. Hence, a neuron $i$ receives a chemical synapse from a neuron $j$ only if $x_j$ is larger than $\theta$}.  {The value of the prefactor $(x_i - \mu_s)$ in Eq. \ref{eq:chem} controls whether the synapses are inhibitory or excitatory. In particular, we chose $\mu_s$ such as $(x_i - \mu_s) < 0$, that is inhibitory chemical synapses}. The numerical values of all parameters are given in Table \ref{tablee}\cite{nobukawa}. %\comment{I would keep the "j" subindex in Eq.~(5).}

Throughout this study, we use symmetric $A^e$ matrices and asymmetric $A^c$ matrices, because the electrical coupling is symmetric, but the chemical coupling is directional. Here we only focus on a proof-of-concept demonstration of the UKF's ability to infer coupling topology; in the future, we plan to study the more challenging (and realistic) scenario of heterogeneous, excitatory, or inhibitory chemical synapses.

\begingroup

\endgroup
\begin{table*}
\caption{\label{tablee} Dimensionless parameters\cite{izke1} used in the simulations of the IM.} %Eq \eqref{izk} for both uncoupled and coupled dynamics.}
\begin{ruledtabular}
\begin{tabular}{cccccccccccccc}
 $a$ & $b$ & $c$ & $d$ & $I\,(I_s)$ & $g_e$ & $g_c$ & $\mu_s$ & $\epsilon$ & $\theta$ & $\alpha$ & $\omega$ & $\sigma_Z$ & $\sigma_{\nu}$  \\ \hline
 $0.2$&$2$&$-56$ &$-16$&$-99$ & $(0,0.1)$ & $(0,0.05)$ & $35$ & $7$ & $0$ & $3$ & $0.15$ & $0.025$ & $0.15$ \\
\end{tabular}
\end{ruledtabular}
\end{table*}

\begin{table*}
\caption{\label{tablee2} Initial guesses of the parameters used by the UKF.} 
\begin{ruledtabular}
\begin{tabular}{cccccccccccccc}
 $a$ & $b$ & $c$ & $d$ & $I\,(I_s)$ & $g_e$ & $g_c$  & $\alpha$ & $\omega$ & $\sigma_P$ \\ \hline
 $(0.01,0.9)$&$(0.01,5)$&$(-70,-40)$ &$(-25,-5)$&$(-94,-104)$ & $(0,0.1)$ & $(0,0.05)$  & $(1,5)$ & $(0.1,1)$ & $0.01$ \\
\end{tabular}
\end{ruledtabular}
\end{table*}

\subsection{The Unscented Kalman Filter}

%The Kalman Filter was introduced by Kalman\cite{kf1} and is used in many \remove{areas}\ajpr{fields} due to its capacity \ajpr{to estimate uncertain random variables which specify the state of dynamical systems.}\remove{of improving the estimation of unknown random variables}. To do so, when considering the evolution of a system, the algorithm takes into account a dynamical model and a sequence of measurements.

%Since the system under study is nonlinear, and the original KF was designed to deal with linear systems, we \remove{turn to}\ajpr{use} the Unscented Kalman Filter (UKF), an alternative better suited to nonlinear estimation \cite{kf2,kf4}. 
% CRIS: YA DICHO EN INTRODUCCION

 {The Kalman filter makes a prediction for the future state of a system, resulting from the state evolution of a dynamical model, and then corrects it using the information coming from experimental data. Even though it was originally developed for linear systems, soon it was extended to include nonlinearities. Different nonlinear extensions were created. The UKF is one nonlinear version of the filter which has a good performance in terms of computational effort.} 

Following the notation in {Forero et al.}~\cite{forero}, we consider the extended state $\bar{\mathbf{u}}$ as the vector given by the state variables $x_i$ and $y_i$ of the $N$ neurons $(i=1,...,N)$ and all the parameters we want to retrieve. Our process model to be employed in the UKF will be $\bar{\mathbf{u}}_{k+1} = \bar{\mathbf{a}}(\bar{\mathbf{u}}_{k})$, where  {$k$ is the timestep index.} $\bar{\mathbf{a}}$ is given by the deterministic part of Eq. \eqref{izk} for the state variables and is the identity operator for the parameters, as we assume they are constant. 
 {In the UKF algorithm, the estimation for $\bar{\mathbf{u}}_{k+1}$ predicted using the stochastic dynamical model is corrected by an experimental piece of data. However, this experimental data will necessarily have some uncertainty resulting from the measurement process, represented by a measurement function.
O}ur measurement function is a selection of the state variables from the extended vector $\bar{\mathbf{u}}_k$, which are perturbed by the measurement noise with standard deviation $\sigma_{\nu}$: $x_i \rightarrow x_i + \sigma_\nu\,\chi^x_i$ and $y_i \rightarrow y_i + \sigma_\nu\,\chi^y_i$, where $\chi$ represents Gaussian white noise. Thus, the covariance matrix of the measurement noise will be $\bar Q_\nu=\sigma_\nu^2\mathbb{I}$. The covariance of the estimated state is $P = \sigma_P\mathbb{I}$, which is evolved by the UKF algorithm from the initial values given in Table~\ref{tablee2}.

\subsection{Implementation}
 
To generate {the} synthetic data that we use as experimental observations, we numerically solve Eq.~\eqref{izk} with a fourth-order Runge-Kutta method, an integration step of $dt = 0.01$, and the parameters reported in Table~\ref{tablee}, keeping the measurements with sampling rate equal to the integration step. With these parameters single (uncoupled) neurons display chaotic dynamics\cite{nobukawa}, as depicted in Fig.~\ref{fig:izkts}(a). Initial conditions for the simulations were drawn from a normal distribution centered at a fixed point of Eq.~\eqref{izk} in the case of no coupling, $(-56.25,-112.5)$, with a standard deviation equal to $1$.

Throughout the study, we employ the UKF implemented by the Python package \textit{FilterPy}\cite{kf3}. The confidence in the process model ($\bar{Q}_Z$) and the measurements ($\bar{Q}_\nu$) are kept constant (see Table \ref{tablee}). An initial transient of $50000$ timesteps was discarded in all runs.

 {The UKF requires an initial guess for the parameters that we want to estimate. To test} the robustness of the UKF, we consider different initial guesses for each run, which are selected from a uniform distribution in the ranges given in Table~\ref{tablee2}. %The choice of a neighborhood for the initial guesses is based on comparing the dynamics that the IM displays with the models available in the literature \cite{izke1,nobukawa}.
%CRIS: NO ENTIENDO ESTA FRASE

To quantify the performance of the UKF in recovering the adjacency matrix $g_e A^{KF} = G_e^{KF}$, we use the Euclidean distance between the original and the recovered matrix:

\begin{equation}
    D(G, G^{KF}) = \sqrt{\sum_{i,j} (G_{i,j} - G_{i,j}^{KF})^2} \ . 
\end{equation}

We quantify the performance using the full coupling matrix $G$, as we want to test not only if the UKF is able to reconstruct the connectivity, but also if it can devise the correct coupling strength without being informed that the coupling strength is the same for all links.  {Also, we chose to use the Euclidean distance because it is a simple, straight-forward measure to compare two graphs of weighted links with a single figure}.

\section{Results}
\subsection{ {Estimation of the parameters of a single neuron}}

First, we illustrate the effectiveness of the UKF in estimating the parameters of a single Izhikevich neuron. The parameters that we attempt to estimate are $a, b, c, d,$ and $I$. Note that the parameters $c$ and $d$ only appear in the resetting dynamics, therefore the UKF can only update them in the event of a spike.
Moreover, the equation for $\dot{y}$ contains a product $ab$, which can increase the uncertainty of the estimation. For example, an underestimation of $a$ can compensate an overestimation of $b$. To avoid such problems, we estimated $ab$ and $a$ independently.

\begin{figure} [t] %fig. 1
    \centering
    \includegraphics[width=0.45\textwidth]{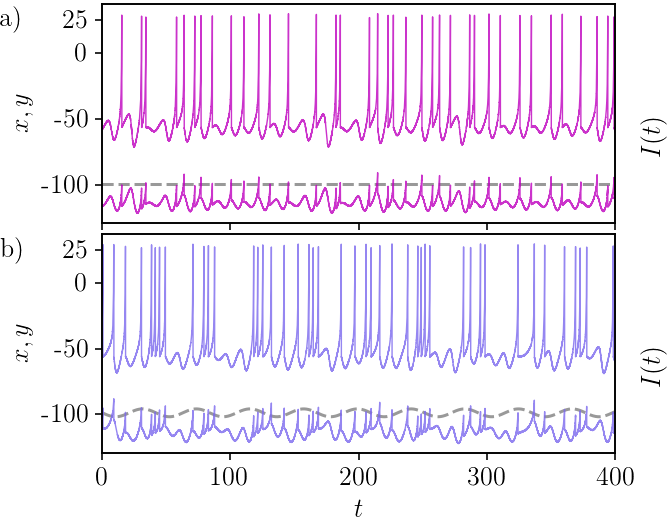}
    \caption{(a) Time evolution of the  variables $x(t)$ (upper curve) and $y(t)$ (lower curve) of an isolated Izhikevich neuron, simulated with Eq. \eqref{izk} with parameters given in Table \ref{tablee}. (b) Response to an external modulated current $I(t) = I_s  + \alpha\sin(\omega t)$. In each panel, the current input is represented by the dashed line. The values of the parameters are given in Table \ref{tablee}.}
    \label{fig:izkts}
\end{figure}

\begin{figure}[t] % fig 2
\includegraphics[width=0.4\textwidth]{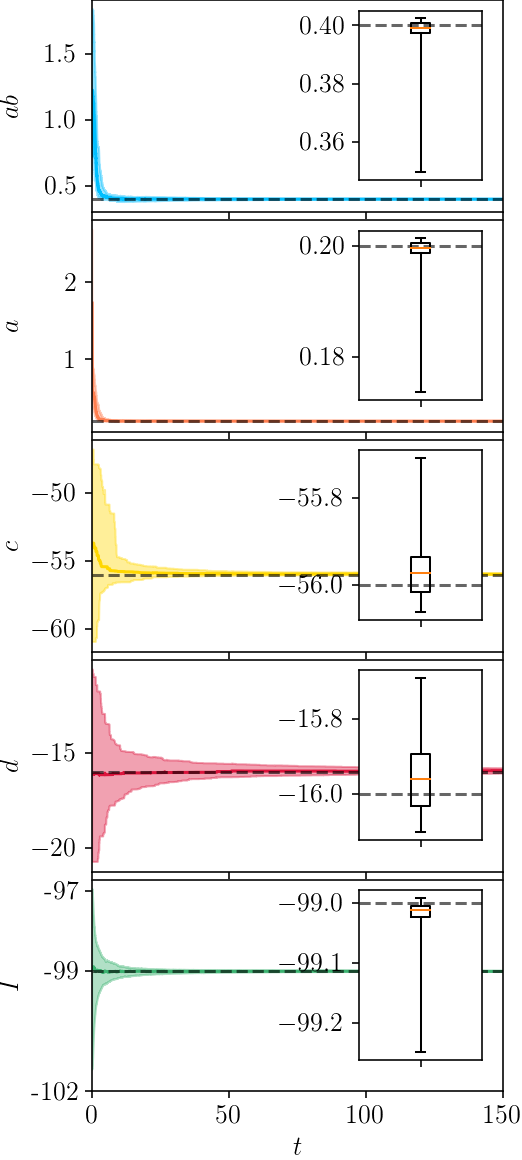}% Here is how to import EPS art
\caption{\label{fig:isoa} Parameter estimation for a single neuron as a function of the simulation time. The colored thick lines represent the median of the estimations computed from 100 runs. The shaded regions represent the first and third quartiles. The dashed lines mark the true values of the parameters. The inset in each subplot shows the distribution of the final estimations. The orange line is the median, the box marks the first and third quartiles and the upper and lower whiskers of the bars represent the maximum and minimum values. The parameter values are given in Table~\ref{tablee}.}
\end{figure}

 {To recover the unknown parameters, we consider $100$ simulated time series as input, each with a different initial parameter guess drawn uniformly from the intervals reported in Table \ref{tablee2}. These intervals have been chosen because in those ranges the spiking of the neuron will be chaotic, which is a piece of information we can infer from the spike sequence}. Results are shown in Fig. \ref{fig:isoa}. For all cases, the real value of the parameter is within the range of the standard deviation. As the estimation of $c$ and $d$ is only updated when the neuron spikes,  {the duration of the simulated time series required to obtain a reliable estimation is larger than for the other parameters.} 

We point out that using the UKF to estimate $c$ and $d$ is unnecessary because a direct estimation of these parameters can be done easily by checking the values of $x$ and $y$ after a spike.

Now, we test the estimation of time-varying parameters. Specifically, we consider a sinusoidal external current, $I(t) = I_s  + 3 \alpha\sin(\omega t)$, and estimate $a$, $b$ and $I(t)$ {(wrongly assuming that the current is constant)}. The effect of such a current on the neuron's dynamics is shown in Fig.~\ref{fig:izkts}(b), where we see that bursts of spikes are followed by periods of subthreshold oscillations. Since at constant $I$ the Izhikevich model displays a great variety of dynamical behaviors including bursting~\cite{izke1}, the inspection of the time series does not provide evidence of the presence of a sinusoidal input current.

\begin{figure} % FIG 3
\includegraphics[width=0.4\textwidth]{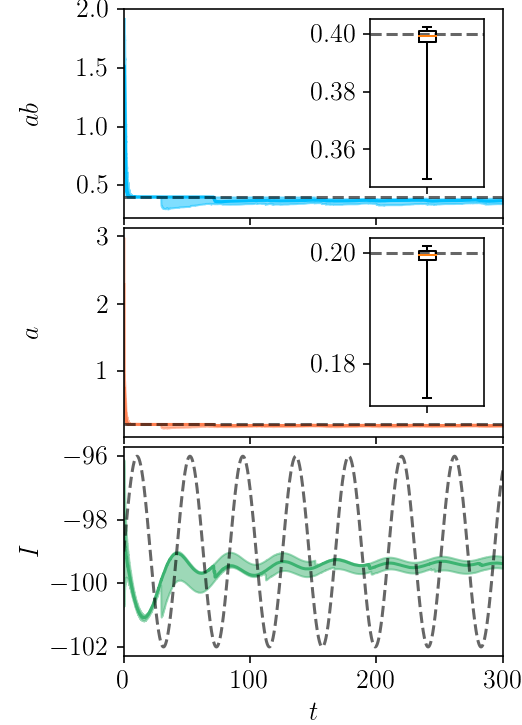}% Here is how to import EPS art
\caption{\label{fig:isof}  Parameter estimation for a single neuron with a time-dependent external current, which is modeled as a constant input. The colored thick lines represent the median of the
estimations computed from $100$ runs. The shaded regions represent
the first and third quartiles. The dashed lines mark the true values of
the parameters. The inset in each subplot shows the distribution of
the final estimations. The orange line is the median, the box marks
the  {first and third quartiles} and the upper and lower whiskers of the
bars represent the maximum and minimum values. The parameter
values are given in Table \ref{tablee}.}
\end{figure}

 {The results of the parameter estimation are shown in Fig.~\ref{fig:isof}. The recovered values of $a$ and $b$ are comparable to those obtained in the previous parameter estimation (see Fig.~\ref{fig:isoa}). The estimated value of $I$ oscillates with a frequency equal to $\omega$, suggesting that $I$ is not constant.} 

\begin{figure} % fig 4
\includegraphics[width=0.4\textwidth]{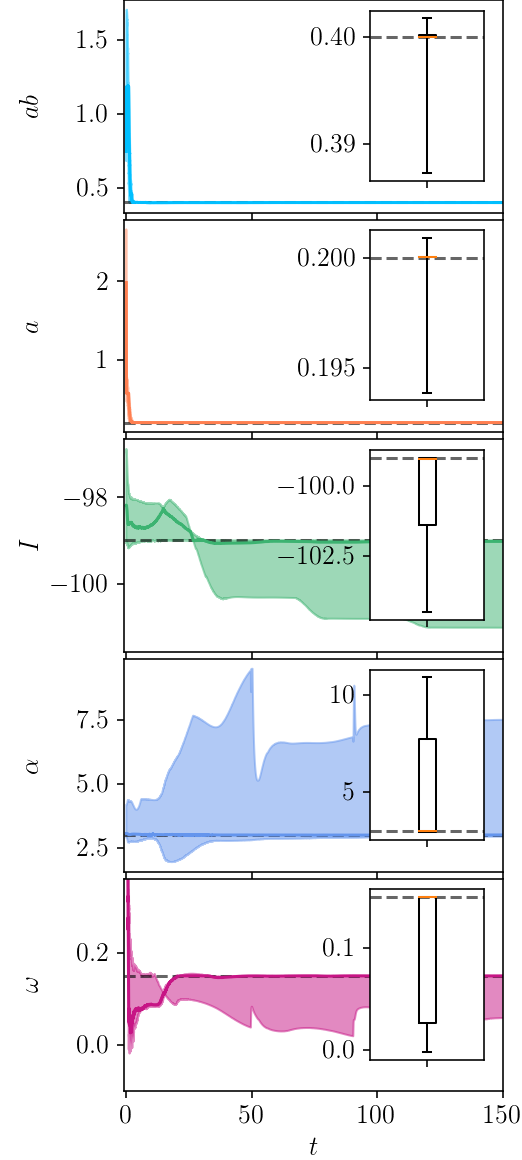}% Here is how to import EPS art
\caption{\label{fig:isof2}  {As Fig. \ref{fig:isof}, but explicitly modeling the input current as $I = I_s + \alpha \sin(\omega t)$. The colored thick lines represent the median of the
estimations computed from $100$ runs. The shaded regions represent
the first and third quartiles. The dashed lines mark the true values of
the parameters. The inset in each subplot shows the distribution of
the final estimations. The orange line is the median, the box marks
the first and third quartiles and the upper and lower whiskers of the
bars represent the maximum and minimum values. The parameter
values are given in Table \ref{tablee}.}}
\end{figure}

Next, we substitute the expression of $I(t)$ in the model, Eq. \eqref{izk}, and separately estimate $I_s$, $\alpha$ and $\omega$. In this case, we also need to include time as an additional dimension of the extended vector space, with dynamic equation $\dot t = 1$. The results of this approach are shown in Fig. \ref{fig:isof2}. The UKF can estimate the correct parameters of the oscillation in the majority of cases. However, large departures from the correct values can be observed, which could be due to the fact that the model with constant $I$ can produce similar output dynamics. 

\subsection{ {Estimation of network connectivity}}

We now consider small networks of Izhikevich neurons, and we investigate the capacity to recover the adjacency matrix $A$ assuming that the coupling strengths, $g_e$ and $g_c$, and
all the internal parameters are known.  {Of course, this is not possible in experiments, and we use these assumptions as a first step for testing the neural network reconstruction problem using the UKF approach: if, given these assumptions, the network cannot be inferred, we can conclude that the UKF approach is not useful; on the other hand, if we succeed in reconstructing the network with these assumptions, as a next step we will test the UKF approach having less information, for instance, assuming a different neuron model, unknown internal parameters, unknown coupling strengths}. We run the UKF algorithm, considering each element of $A$ as an additional dimension of the extended vector $\bar{\mathbf u}$.

We consider networks with $N = 4$ neurons, which can be seen as building blocks of bigger networks.  {However, we must keep in mind that complex systems usually display emergent collective behaviour when the number of elements is large enough, and therefore, while the UKF algorithm may succeed in reconstructing the topology of a small network, the collective behaviour that may emerge for a large enough number of neurons (and/or the large number of parameters to be inferred), will probably cause the UKF algorithm to fail. Therefore, the study of the role of the network size is of course important and additional work is planned, that will be reported elsewhere.}

The
network topologies are shown in Appendix A, see Figs. \ref{fig:ap1} and \ref{fig:ap2}. The simulated time evolution of the
membrane potentials of all nodes for each network topology are also shown in Figs. \ref{fig:ap1} and \ref{fig:ap2}. The network
dynamics differ in their level of synchronization. 
First, we consider electrical coupling  {only ($g_c = 0$)}, such that the adjacency matrix is symmetric and we only need to determine its upper triangular elements $G_e = g_e A^e$ with $g_e = 0.05$. 

\begin{figure}[h] %Fig 5
    \centering
    \includegraphics[width=0.4\textwidth]{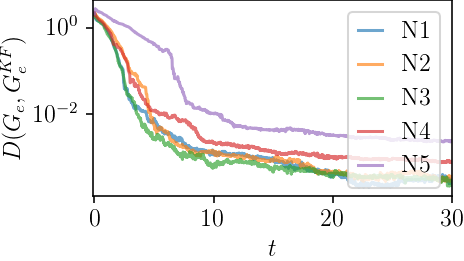}
    \caption{Evolution of the distance $D$, which represents the Euclidean distance between real coupling matrix $G_e$ and the estimated one $G_e^{KF}$. N1, N2, N3, N4, and N5 represent different topologies, as shown in Fig.~\ref{fig:ap1}. For all topologies $D$ {decreases sharply at first, then it saturates below $10^{-2}$.} The coupling strength is $g_e = 0.05$ and the dynamics displayed by the networks are shown in Fig. \ref{fig:ap1}. 
    }
    \label{fig:eee1}
\end{figure}

The results for the different topologies are displayed in Fig.~\ref{fig:eee1}, where we see that the UKF gives an excellent estimation of $G_e$, as the Euclidean distance $D(G_e, G_e^{KF})$ approaches $0$.

To study the effect of chemical synaptic coupling in the UKF estimation, we add direct links between some nodes in the formerly symmetric networks, as shown in Appendix A, Fig. \ref{fig:ap1}. 
We estimate two adjacency matrices, one encoding electrical coupling, $g_e A^e = G_e$ with $A^e = (A^e)^T$, and the other encoding chemical coupling, $g_c A^c = G_c$ with $A^c \neq (A^c)^T$. {We chose $g_e = 0.1$ and $g_c = 0.05$. Note that we increase $g_e$ compared to the previous case to test the robustness of the UKF against synchronized states.} Since we use inhibitory synapses, the firing rate decreases slightly. In the limit of total synchronization the input through electric coupling, $E_i$ goes to zero, while $C_i$ is exactly the same for each element in the network.
%The dynamics displayed by the networks are shown in Fig. \ref{fig:apx2}. Again, depending on the network architecture, the nodes present different levels of synchronization.

%{\bf{CRIS: CAN WE INCLUDE ONE SHORT COMMENT ON WHICH IS THE EFFECT OF THESE CHEMICAL LINKS ON THE NETWORK DYNAMICS? WHAT HAPPENS IF $g_c = 0.0$? }}

\begin{figure} % fIG 6
    \centering
    \includegraphics[width=0.4\textwidth]{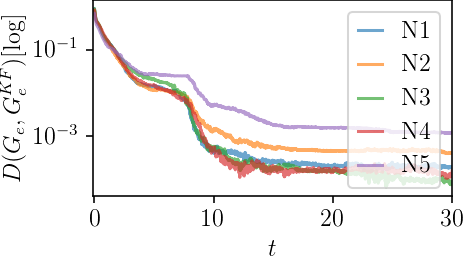} \\
    \includegraphics[width=0.4\textwidth]{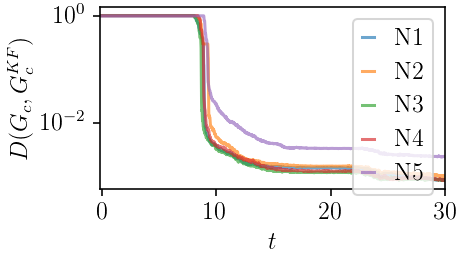}
    \caption{Evolution of the distance $D$ for (a) electrical coupling and (b) chemical coupling, with coupling strengths $g_e = 0.1$ and $g_c = 0.05$. The distance between the original and the estimated adjacency matrices, $D(G, G^{KF})$, decreases sharply with simulation time, saturating below $10^{-2}$ for all network topologies. The dynamics displayed by the networks are shown in Fig. \ref{fig:ap2}.}
    \label{fig:ebe2}
\end{figure}

We see in Figs.~\ref{fig:ebe2}(a) and \ref{fig:ebe2}(b) that even with two coupling schemes, the chemical one being nonlinear, the UKF can estimate the correct coupling matrices.  {All networks are heterogeneous} in the sense that the number of connections is not the same for the different neurons. As pointed out by {Forero et al.}~\cite{forero}, the UKF is robust against synchronization, which is confirmed here.

 {Here we presented reconstruction results in the case of inhibitory synapses, however, we checked that the UKF provide similar results also for excitatory synapses and a mix of excitatory and inhibitory synapses, provided it knows which synapses are excitatory and which are inhibitory.}

We highlight that for all cases the Euclidean distance $D(G,G^{KF})$ saturates  {below} $10^{-2}$. Furthermore, to verify that all the links were correctly estimated we classified the performance of the UKF using the Receiver Operating Characteristic  (ROC) curve\cite{fawcett}.  {If the UKF recovers the right connectivity, then the Area Under the ROC Curve (AUC)\cite{fawcett} will be 1.} For all cases studied, we obtained an AUC $> 0.99$, implying a perfect reconstruction of the underlying topologies,  {that is, the UKF predicts a link between two neurons $i$ and $j$ only if $A_{ij}=1$}.  {We believe that the UKF is robust against noise as long as noise can be seen as a small perturbation to the system and the dynamics is not driven by it.}

%since $D(G,G^{KF})$ tends to zero in all cases. {\bf{CRIS: GOES TO ZERO? REALLY? THIS IS NOT SEEN IN THE FIG.}}

%\begin{figure}
%    \centering
%    \includegraphics[width=0.4\textwidth]{euc_ebc100q.png}
%    \caption{As Fig. \ref{fig:ebe2} but for the chemical adjacency matrix.}
%    \label{fig:ebc2}
%\end{figure}

\subsection{ {Estimation of network connectivity in temporal networks}}

Finally, we consider temporal networks, in which $G_{e,ij}=g_e~A_{ij}$ varies with time. This is the case in many applications of network theory~\cite{holme}, and in neuroscience, it is especially important since it can be linked to plasticity~\cite{citri}.

We model time-varying networks by considering  {couplings between neurons that switch on at a simulation time $t =200$. More precisely, three single neurons connect at $t = 200$} in a linear chain, ($1\leftrightarrow 2 \leftrightarrow3$),
\begin{equation}
    g_e\begin{bmatrix}
    0 & 0 & 0 \\
    0 & 0 & 0 \\
    0 & 0 & 0
    \end{bmatrix}
    \rightarrow
    g_e\begin{bmatrix}
    0 & 1 & 0 \\
    1 & 0 & 1 \\
    0 & 1 & 0
    \end{bmatrix} \ .
\end{equation}
We assume that all the internal parameters are known and we only estimate the network's topology. 

\begin{figure} %FIG 7
    \centering
    \includegraphics[width=0.4\textwidth]{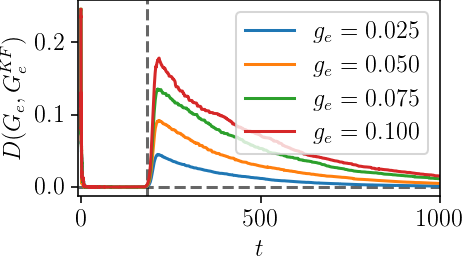}
    \caption{Evolution of the distance $D$ between the adjacency matrix and the estimated one in the case of time-dependent coupling. $D$ is depicted as a function of time, for different coupling strengths. The vertical dashed line represents the instant in which the coupling is turned on and the horizontal one marks the zero.}
    \label{fig:plas}
\end{figure}

The results are presented in Fig.~\ref{fig:plas}.  {Before the coupling is switched on, the UKF has quickly inferred the absence of coupling (as $D\rightarrow 0$). After the coupling is switched on, $D$ first increases sharply and then decreases steadily for all values of $g_e$.} This means that the UKF can detect the emergence of coupling and estimate the $G_e$ matrix correctly. However, the estimation after the change in the network takes more time than the initial estimation of the null adjacency matrix. This is because the covariance on the matrix coefficients will decrease, meaning high confidence in the inferred matrix before the coupling is switched. When the matrix is changed, the filter has to adjust to the new state, but the low covariance will make the convergence rate slow. Nevertheless, the filter is eventually able to recover the right network structure. Different initial model or state covariances are expected to impact the convergence time, both before and after turning on the coupling.  {Higher covariances will result in higher variability in the predictions. This variability is more efficient in capturing changes in the parameters. On the contrary, when covariances are smaller, the predictions are less prone to change and thus adapt to new values. The right choice of this parameter will result in a responsive system with sufficiently stable inferred parameters.}

\section{Discussion and Conclusions}

We studied the capability of the UKF for recovering the parameters of a single neuron and of small neural ensembles modeled with the Izhikevich model. We simulated the equations governing the system dynamics and used the simulated time series as experimental observations to feed the UKF algorithm, with confidence regulated by $\bar{Q}_\nu$. The IM was the process model with confidence regulated by $\bar{Q}_Z$.

When the parameters of an isolated neuron are constant in time, the UKF is able to estimate all the parameters. Second, we studied an isolated neuron with a sinusoidal input current, which displayed bursting spike dynamics.  Due to the rich variety of dynamical behaviors of the IM, it is not trivial to identify the cause of the bursting activity. Still, even when modeling the current as a constant, the UKF retrieved the neuron parameters and the average value of the current and suggested an oscillating current. When including the oscillating current in the process model, the UKF was able to provide a reasonable estimate of the  {amplitude ($\alpha$), the mean ($I$), and frequency of the oscillation ($\omega$).}

We have also estimated the connectivity of small networks of Izhikevich neurons  {with known internal parameters.} First, we analyzed the five possible network topologies for four neurons with undirected electrical coupling. Then, we added directed chemical connections to the same networks. The UKF was able to recover the connectivity for all the networks regardless of the synchronization level. %between neurons, which we quantified with the Kuramoto order parameter. 

Finally, we addressed the problem of temporal networks by analyzing a network of three electrically coupled neurons, in which the topology changed from no coupling to a chain topology. The UKF was able to identify the change in the network and estimate the connectivity correctly

The results presented here were obtained considering measurements of both $x$ and $y$. Beyond that, we conducted a preliminary analysis of the applicability of the UKF when only measurements of the $x$ variable are available. Our results suggest that the UKF is still able to recover the parameters of a single neuron and the network connectivity. However, to obtain good estimates, the UKF hyperparameters had to be carefully tuned, in particular, the standard deviation $\sigma_\nu$ and the initial condition for $\sigma_P$. %Is also important to track the robustness of UKF against the sampling of the data points. 

As in experimental measurements  {only short time series with limited temporal resolution can be recorded,} further work is needed to clarify the impact of the duration of the time series and the sampling time. While the results presented here were obtained using each simulated data point (i.e, using the integration step as sampling time), preliminary studies suggest that the UKF is robust to downsampling up to 1:20, if the time series is long enough.

 {Future work should also address larger networks and different types of neurons. In fact, as discussed before, complex systems usually display emergent collective behavior when the number of elements is large enough. Therefore, the UKF algorithm may succeed in reconstructing the topology of a small network, but will probably fail for a large number of neurons, or when there is a large number of unknown parameters. Therefore, further work is planned to test the UKF algorithm when the networks are larger and when the internal and coupling parameters are unknown. While we expect that the UKF algorithm will fail to reconstruct the network, it may yield some information that can be useful for inferring some properties of the real network (e.g., the average degree, the degree distribution, the modularity, etc.).}

 {Finally, it will be interesting to check if the UKF can differentiate between inhibitory and excitatory synapses.}

\begin{acknowledgments}

R.P.A. acknowledges financial support from Coordenação de Aperfeiçoamento de Pessoal de Nível Superior–Brasil (CAPES), Finance Code 001. H. A. C. thanks ICTP-SAIFR and FAPESP grant 2021/14335-0. G.T. and C.M. acknowledge the support of the ICREA ACADEMIA program of Generalitat de Catalunya and Ministerio de Ciencia e Innovación, Spain, project PID2021-123994NB-C21.

\end{acknowledgments}

\appendix

\begin{figure}[t] %fig8
\includegraphics[width=\columnwidth]{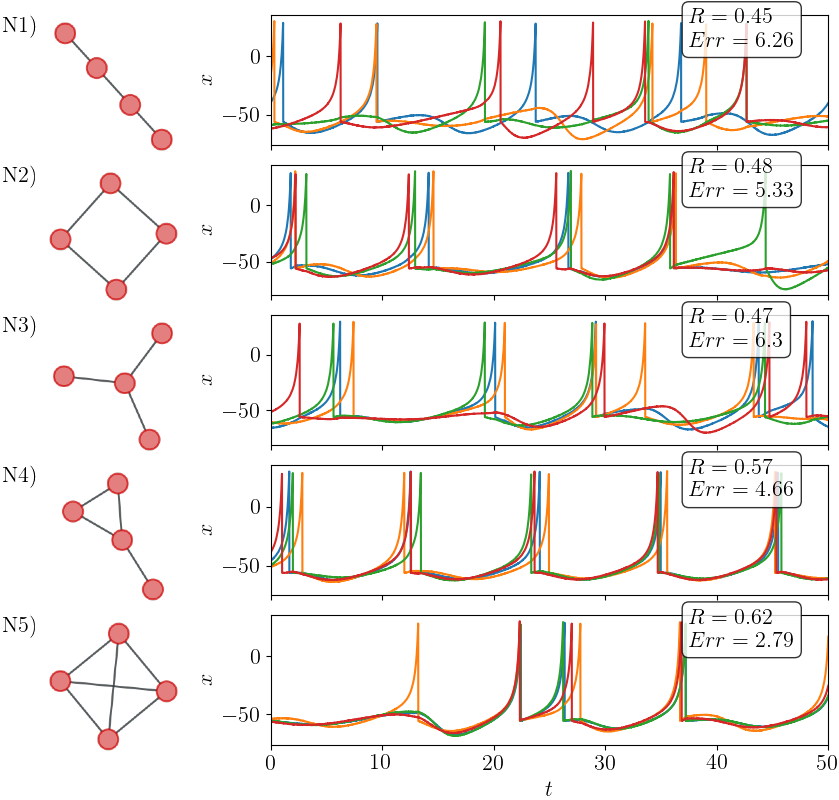}% Here is how to import EPS art
\caption{\label{fig:ap1} The five network topologies used when considering only electrical coupling (Eqs. \eqref{izk} and \eqref{eq:elec}).  {The links between nodes are represented by black lines. %Note that cases b) and e) are examples of homogeneous networks, while the others are heterogeneous networks. 
For each case, we show the network's dynamics with $g_e = 0.05$.} The insets show the Kuramoto order parameter $R$ and the synchronization error $Err$. }
\end{figure}

\begin{figure}[t] %fig9
\includegraphics[width=\columnwidth]{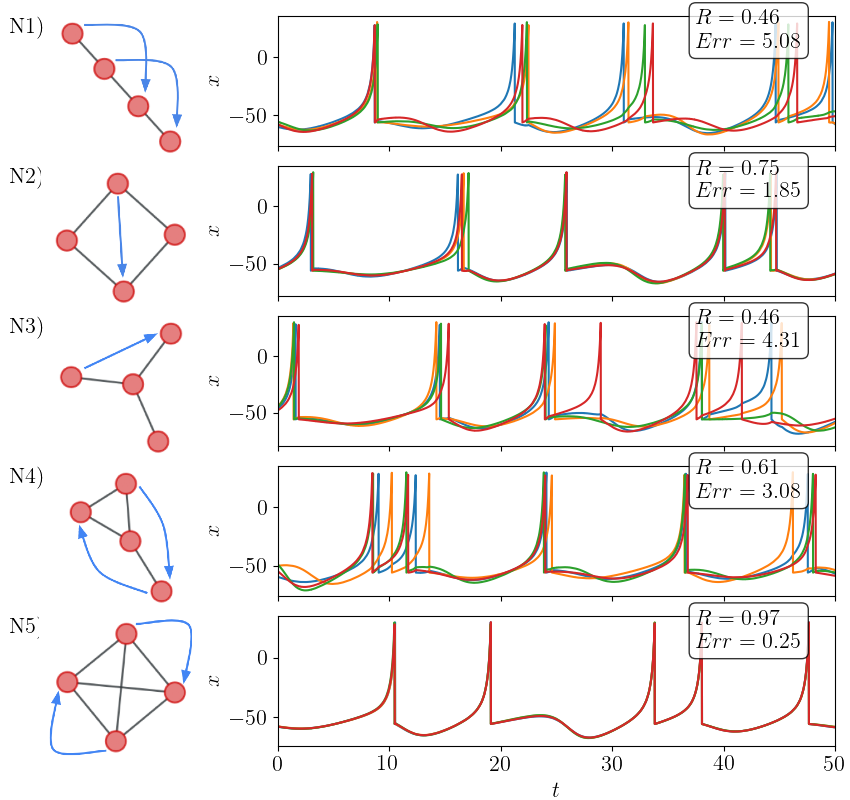}% Here is how to import EPS art
\caption{The five network topologies used when considering both electrical and chemical coupling (Eqs. \eqref{izk}, \eqref{eq:elec}, and \eqref{eq:chem}), the undirected links are electrical, and the directed links are chemical.
The left column displays the network's dynamics with $g_e = 0.10$ and $g_c = 0.05$. The insets show the Kuramoto order parameter $R$ and the synchronization error $Err$.}
\label{fig:ap2}
\end{figure}

\section{Topologies and synchronization quantification}

The network topologies considered when studying electrical coupling between nodes are presented in Fig.~\ref{fig:ap1}. The black links represent undirected connections between nodes, so the resulting adjacency matrices are symmetric ($A = A^T$). The simulated time evolution of the membrane potentials of all nodes for each network topology is also shown in Fig. \ref{fig:ap1} on the right side, 
together with two synchronization measures, the Kuramoto order parameter~\cite{kuramoto}{$R$} and the synchronization error $Err$.

To evaluate the Kuramoto order parameter, we assign a phase $\phi$ for each neuron time series that grows linearly at each spike with a gain of $2\pi$ as defined in {Ivanchenko et al.}~\cite{ivanchenko}. The Kuramoto order parameter is given by
\begin{equation}
    R = \frac{\big<|\sum_{j=1}^{N} e^{ i\phi_j(t)}|\big>_t}{N} \ ,
\end{equation}
where $N$ is the number of oscillators considered in the measure and the average is taken over time. For totally synchronized systems, $R = 1$. For totally unsynchronized systems, $R \approx 0$.  

Likewise, the synchronization error gives us an idea of how synchronized the system is, we apply it directly to the time series. First, we calculate the average membrane potential $\bar{x}$ of all oscillators in the network. Then, we compute how much each oscillator deviates from $\bar{x}$. Thus, the synchronization error is computed as
\begin{equation}
    Err = \Big<\frac{\sum_{i=1}^N |x_i(t) - \bar{x}(t)|}{N}\Big>_t \ .
\end{equation}
Hence, $Err = 0$ in the case of total synchronization, where $x_i = x_j $, $\forall~(i,j) \in [1,N]$. While for unsynchronized systems, $Err$ may assume large values.

When both electrical and chemical coupling between nodes are considered, we use the topologies presented in Fig. \ref{fig:ap2}. The adjacency matrices are not symmetric ($A \neq A^T$), and all the networks are heterogeneous, meaning that the nodes have a different number of connections.  {The simulated time evolution of the membrane potentials of all nodes for each network topology is also shown in Fig. \ref{fig:ap2} on the right side.}

%\newpage

\bibliography{references}% Produces the bibliography via BibTeX.

\end{document}